\documentclass[manuscript]{aastex61}
\usepackage[utf8x]{inputenc}
\usepackage{graphicx}
\usepackage{textcomp}
\usepackage{bm}
\usepackage{amsmath}
\usepackage{gensymb}
\shorttitle{Untwisting magnetic field}
\shortauthors{Sangeetha et al.}
\begin{document}
\title{Signatures of Untwisting Magnetic Field in a Small Emerging Bipole in the Solar Photosphere}
\author{C.~R.~Sangeetha}
\affil{Inter-University Centre for Astronomy and Astrophysics, Post Bag-4, Ganeshkhind, Pune 411007, India}
\affil{Udaipur Solar Observatory, Physical Research Laboratory, Badi Road, Udaipur 313001, India}
\email{sangeetha@iucaa.in}

\author{Durgesh Tripathi}
\affil{Inter-University Centre for Astronomy and Astrophysics, Post Bag-4, Ganeshkhind, Pune 411007, India}

\author{S.~P.~Rajaguru}
\affil{Indian Institute of Astrophysics, Bangalore-34, India}
\begin{abstract}

We perform a study of fluid motions and its temporal evolution in and around a small bipolar emerging flux region using observations made by the Helioseismic and Magnetic Imager (HMI) on-board the Solar Dynamics Observatory (SDO). We employ local correlation tracking of the Doppler observations to follow horizontal fluid motions and line-of-sight magnetograms to follow the flux emergence. Changes in vertical vorticity and horizontal divergence are used to derive signatures of evolving twists in the magnetic field. Our analysis reveals that the two polarities of the magnetic flux swirl in opposite directions in early stages of flux emergence indicating an unwinding of the pre-emergence twists in the magnetic field. We further find that during the emergence, there is an increase in swirly motions in the neighbouring non-magnetic regions. We estimate the magnetic and kinetic energies and find that magnetic energy is about a factor of ten larger than the kinetic energy. During the evolution, when the magnetic energy decreases, an increase in the kinetic energy is observed indicating transfer of energy from the unwinding of magnetic flux tube to the surrounding fluid motions. Our results thus demonstrate the presence of pre-emergence twists in emerging magnetic field that is important in the context of the hemispheric helicity rule warranting a detailed statistical study in this context. Further, our observations point to a possible widespread generation of torsional waves in emerging flux regions due to the untwisting magnetic field with implications for upward energy transport to the corona.

\end{abstract}

\section{Introduction}
The emergence and dynamics of magnetic flux play a pivotal role in the formation and evolution of various structures in the solar atmosphere as well as have a direct impact on space weather and climate. On the one hand, they are considered to be the prime candidates for the transfer of mass and energy within the solar atmosphere \citep[e.g.,][references there in]{2017LRSP...14....2B}. On other hand, they are known to be closely associated with eruptive events at various scales such as flares \citep[e.g.,][references there in]{2018A&A...612A.101V,2017RNAAS...1...24S}, jets \citep{2018ApJ...861..108Z,2016A&A...589A..79M}, coronal mass ejections \citep[CMEs;][]{2017ApJ...845...18Y,2017ApJ...850...95S, 2004A&A...422..337T, 2007A&A...472..967C}, UV bursts \citep{2018ApJ...854..174T,2019ApJ...871...82G,2015ApJ...809...82G}, Ellerman Bombs \citep{2007ApJ...657L..53I,2004ApJ...614.1099P}.

The emergence of magnetic flux occurs at various spatio-temporal scales. Some emerge and evolve into prominent active regions (ARs), whereas some get fragmented during the early phase. ARs, when observed on the solar surface, exhibit a well-known characteristic pattern:  those in the northern hemisphere show counter-clockwise superpenumbral structures, whereas those observed in southern hemisphere show clockwise superpenumbral structures as was first discovered by \citet{1927Natur.119..708H}. This pattern has been well studied since then and is attributed to the helicity of magnetic field, and is now known as the hemispheric helicity rule \citep{1995ApJ...440L.109P,1998ApJ...507..417L,2014ApJ...783L...1L,2014SSRv..186..285P}.

In recent years, with the advent of long-term vector magnetic field measurements statistical studies have been performed to understand the hemispheric helicity rule. \citet{2013ApJ...775L..46W}  has shown that about 60{--}82.5\% of the active regions follow the hemispheric helicity rules. By taking a sample of 151 active regions, \citet{2014ApJ...783L...1L} have found that 75\% $\pm$7\% of the active regions obey the hemispheric helicity rule. The question then arises that what happens to the other 25\% that they do not follow this pattern. We emphasise that \citet{2014ApJ...783L...1L} did not discriminate between newly emerging and fully evolved active regions. In a study of a sample of 28 ARs, which focussed on newly emerging ARs, \citet{2014ApJ...785...13L} found that only 61\% obeyed the hemispheric helicity rules, which is a significant drop in the percentage of active regions following the hemispheric pattern. The authors suggest that this could be due to small sample size and therefore may not be statistically significant. However, such a result also suggests that emerging flux regions may go through complex processes that may prevent them from exhibiting the peculiar behaviour. Therefore, this warrants further detailed studies on emerging flux regions.

In terms of magnetic helicity, the observed hemispheric pattern corresponds to negative (positive) magnetic helicity in the northern (southern) hemisphere \citep{1990SoPh..125..219S,1995ApJ...440L.109P,1997SoPh..174..291A,1998ApJ...496L..43B,2014SSRv..186..285P}. \citet{2000SoPh..192..177D,2003ESASP.517...43G} suggested that the large-scale converging flows towards active regions with the action of Coriolis force are consistent with the observed hemispheric pattern \citep{2014SSRv..186..285P,2019ApJ...873...94B}. Therefore, we may deduce that the magnetic helicity of AR’s showing the hemispheric rule could primarily be due to the swirling fluid, which converts its kinetic helicity to the magnetic helicity. However, the large-scale converging flows develop as a consequence of the thermal imbalance due to the establishment of fully grown sunspots \citep{1986ApJ...306..751W,2002Ap&SS.279..389K}. Therefore, the hemispheric pattern may not be unambiguously observed in newly emerging active regions. The reduced trend in the hemispheric pattern in case of emerging active regions observed by \citet{2014ApJ...785...13L} is consistent with such a picture.  However, we stress that the origin of helicity is still not established, and that the above inferences based on large-scale inflows is only suggestive of a possible origin.


An essential diagnostic of the above-suggested scenario is then to study the signatures of pre-emergence magnetic helicity (twist and writhe) in emerging AR’s. If the sign of pre-emergence twist does not match with that is expected from the hemispheric pattern, then the emerging AR would undergo untwisting motion to follow the hemispheric pattern after complete emergence. Therefore, a careful analysis is warranted to fully comprehend the nature of flux emergence and its interaction with the surrounding fluid. For example, if we study the fluid vorticity during the process of the flux emergence, we may observe signatures of winding or unwinding of flux tubes.

In addition to the above, several studies show that magnetic regions on the Sun are associated with vortex motions \citep{1988Natur.335..238B,2010ApJ...723L.139B,2012Natur.486..505W,2014ApJ...797...52Y,2016ApJ...826....6Z}. These motions are of particular significance because they can generate torsional Alfv$\acute{\text{e}}$n  waves \citep{1972SoPh...27...71G,2016NatPh..12..179J,2018MNRAS.480.2839L,2019A&A...621A..43F}, which are important for solar coronal heating. Due to this reason, a search on these torsional motions is underway \citep{1988Natur.335..238B,1995ApJ...447..419W,2009A&A...493L..13A,2010ApJ...723L.139B,2012Natur.486..505W,2006ApJ...646L..85Z,2013ApJ...772...52G,2016ApJ...824..120S}. However, studies are sparse on the nature of twists observed in the early stages of flux emergence \citep{1996ApJ...462..547L,2003ApJ...593.1217P,2001SoPh..203..289P}. 

For the above purposes, we observe a small bipolar region during its emergence in the photosphere. This region was a part of the data used in the analysis of the work carried out by the two authors of this paper \citet{2016ApJ...824..120S}. The region showed different behaviour than the rest of the regions used for analysis. When investigated on this further, we realised that it was due to the small flux emergence in the southern hemisphere. This motivated us to investigate this region further using continuous observations provided by Helioseismic and Magnetic Imager (HMI) onboard Solar Dynamics Observatory \citep[SDO;][]{2012SoPh..275..207S} to study the changes in vorticity during the emergence. The data used for this analysis is discussed in \S \ref{obs}. The analysis and results are discussed in \S \ref{dar}. Finally, we present the summary and conclude in \S \ref{SandC}. 

\section{Observation and Analysis Method}\label{obs}
\subsection{Data}
\begin{table}[ht]
\centering
\caption{\footnotesize The time series name along with the start and end of observation for each dataset is shown in the table.}
\begin{tabular}{ccc}
\hline
{\bf Time Series}  &  {\bf Start of observation}  &   {\bf End of Observation} \\[0.5ex]
  &   (--:-- UT dd/mm/yyyy)   &   (--:-- UT dd/mm/yyyy)   \\[0.5ex]
\hline
T1  & 02:00 on 08/05/2011  & 16:00 on 08/05/2011  \\[0.5ex]
T2  & 16:00 on 08/05/2011  & 06:00 on 09/05/2011  \\[0.5ex]
T3  & 06:00 on 09/05/2011  & 20:00 on 09/05/2011  \\[0.5ex]
T4  & 20:00 on 09/05/2011  & 08:00 on 10/05/2011  \\[0.5ex]
T5  & 00:00 on 11/05/2011  & 14:00 on 11/05/2011  \\[0.5ex]
\hline
\end{tabular}
\label{tab:data}
\end{table}

\begin{figure*}[ht]  
\centering
\includegraphics[width=0.5\textwidth]{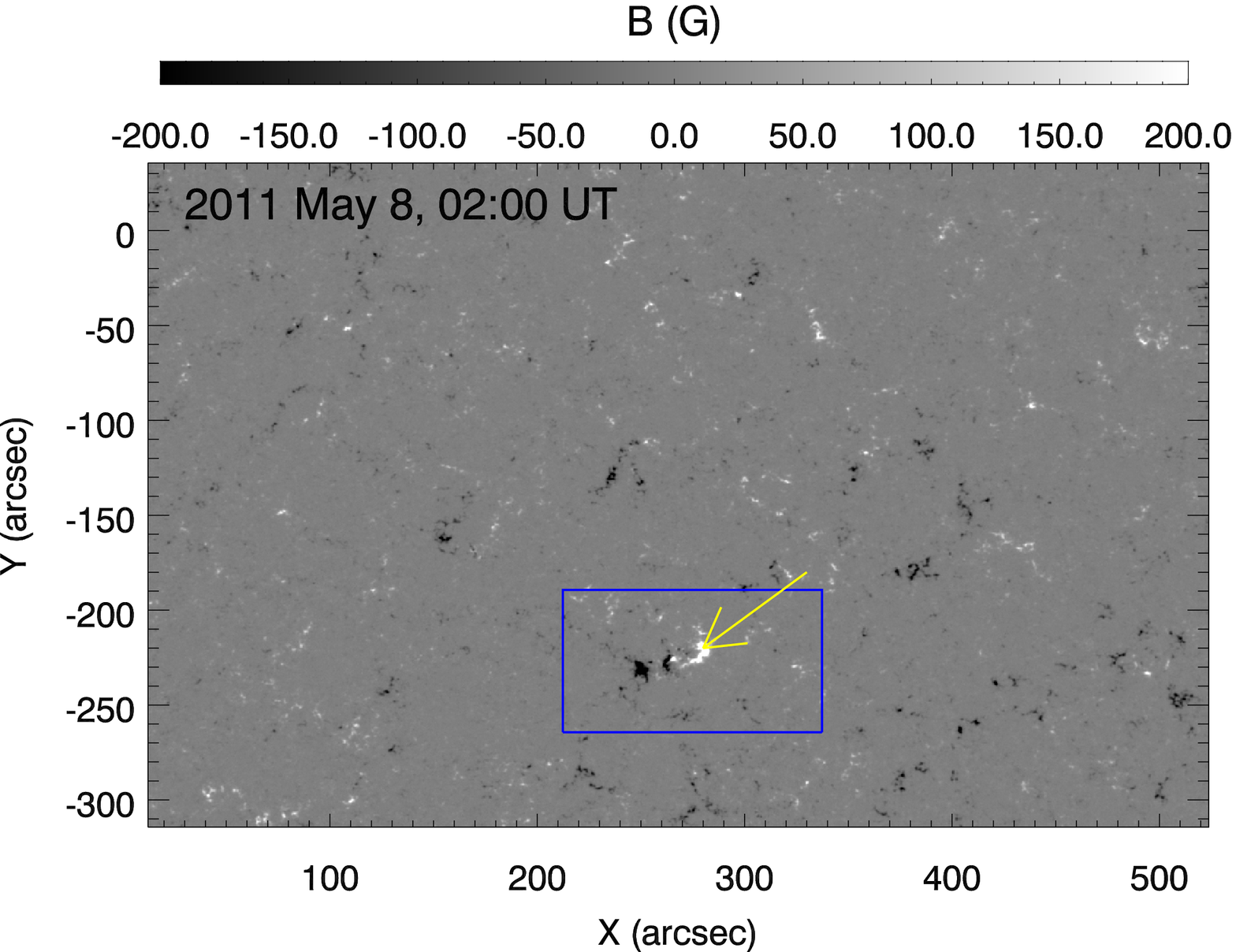}
\caption{LOS magnetic field of the emerging flux taken at 02:00~UT on 2011 May 08. The image is displayed between $\pm$~200~G to enhance the small-scale features. The region enclosed by the over-plotted blue box is considered for further analysis. The yellow arrow indicates the emerging flux region.}\label{mg08}
\end{figure*}

A small bipolar region that is studied in this paper started to emerge on May 8, 2011 in the southern hemisphere of the Sun, and was later identified as \textsl{NOAA~11211} on May 10, 2011.  The focus of this work is to study the evolution of this emerging flux region and its effects on the plasma motions in its vicinity. For this purpose, we have used the Doppler velocity ($v_{d}$), and line-of-sight (LOS) magnetic field ($B_{LOS}$) recorded by the HMI instrument on-board SDO. The HMI provides measurements of these quantities with a cadence of 45~s and pixel size of 0.5$\arcsec$.  The whole region taken for the analysis is 512$\arcsec \times $512$\arcsec$ which is centred around Carrington Longitude 16.1$\degree$ and Carrington latitude -3.5$\degree$. Figure~\ref{mg08} displays the LOS magnetic field maps showing emerging bipole located with a yellow arrow. The over-plotted blue box encloses the region that is considered for further detailed analysis.

We have tracked the marked bipolar region from May 8th till May 11.  To study the evolution of flows and magnetic field, we subdivided the full length of observations into five shorter segments namely T1, T2, T3, T4 and T5 as noted in Table~\ref{tab:data}. Each observation set covers 14 hours, except T4 that includes only 12 hours, due to unavailability of data. In order to avoid large uncertainties due to projection effects, we demanded that the region of interest should be located within 30$\degree$ latitude and longitude. The error due to projection effect would be less than 15\%. Moreover, the duration for each time series is taken so as to be not influenced by the effects of dying and newly generated supergranular flows \citep[see e.g.,][]{2016ApJ...824..120S}.

\begin{figure*}[ht]  
\centering
\includegraphics[width=0.8\textwidth]{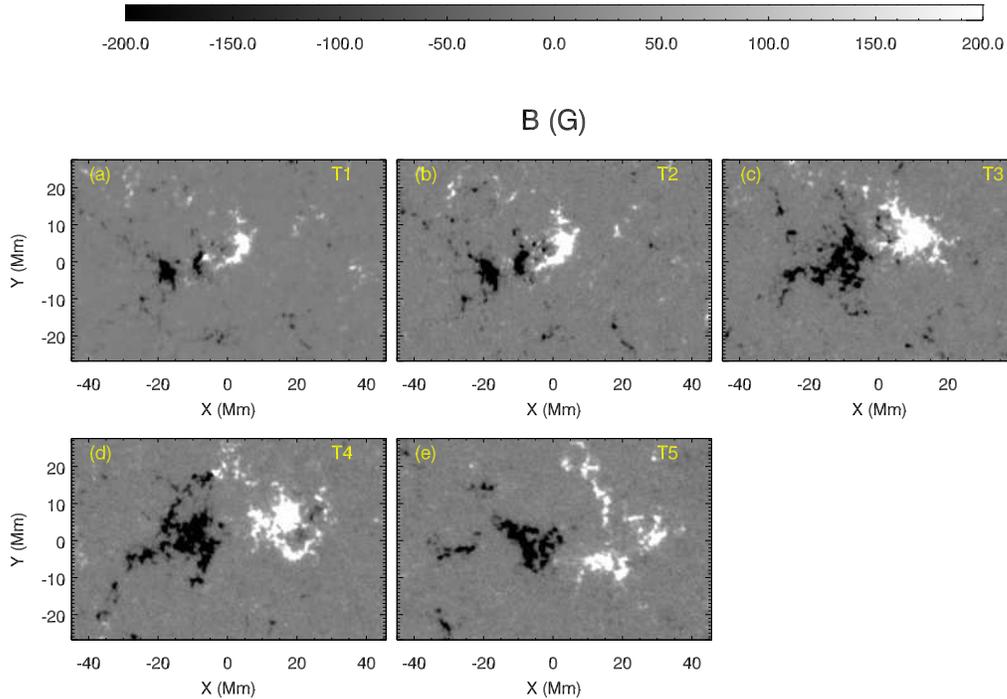}
\caption{First snapshots of LOS magnetic field of the emerging flux plotted for different time series T1, T2, T3, T4 and T5, displayed between $\pm$~200~G. The field of view is same as the blue box seen Figure~\ref{mg08}.} \label{mg_all}
\end{figure*}
\begin{figure*}[ht]  
\centering
\includegraphics[width=0.8\textwidth]{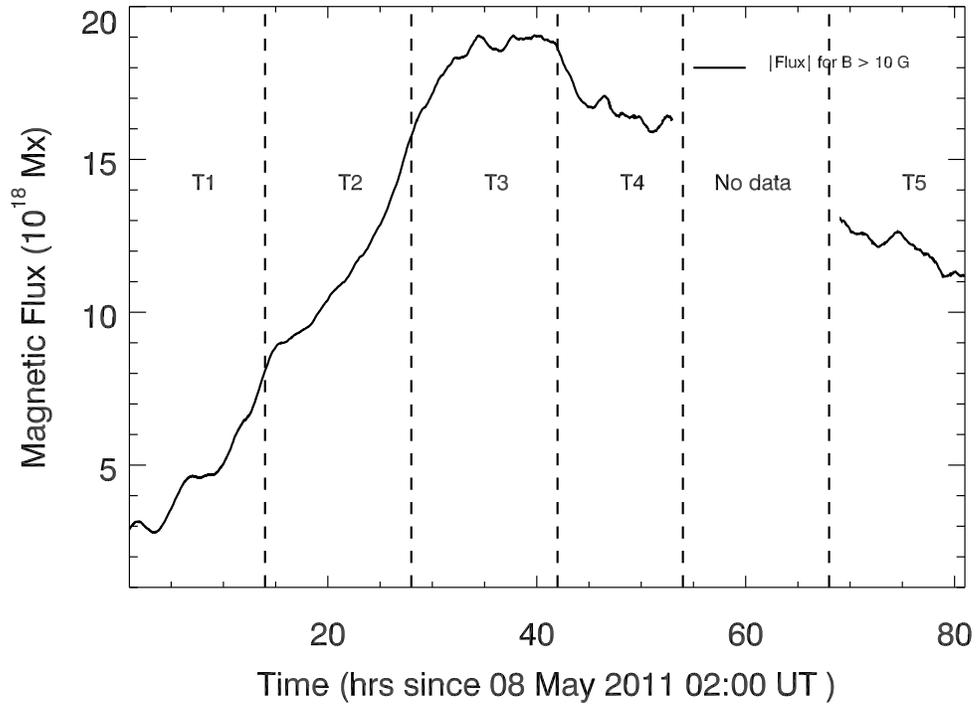}
\caption{Evolution of magnetic flux derived from HMI LOS magnetic field for the emerging flux for the different time sets T1, T2, T3, T4 and T5. } 
\label{mg_flux}
\end{figure*}

Fig.~\ref{mg_all} displays the HMI LOS magnetic field maps corresponding to the blue box in Figure~\ref{mg08} at the beginning of each observation set. We note here that the leading polarity is positive whereas the trailing polarity is negative. As the figure reveals, with passing time, the bipolar region spreads and show enhanced 
magnetic field. In Fig.~\ref{mg_flux}, we plot the evolution of magnetic flux derived within the area indicated by Fig.~\ref{mg_all} for all the observation datasets. Fig.~\ref{mg_flux} reveals that the magnetic flux continuously increases and reaches the highest values of $\sim\ 18\ \times\ 10^{18}$~Mx during the third observation set. 
Thereafter, the flux starts to decrease and continue to decline throughout our observation. The missing data between T4 and T5 marked as 'no data' in the figure. 

\subsection{Analysis}

The main aim of the paper is to study the interplay between the emerging flux and fluid motion. For this purpose, we have derived the vertical vorticity ($\omega_{z}$) and horizontal divergence ($d_{h}$), which are defined as:

\begin{equation}
 \centering
  (\bm{\nabla} \times \bm{v})_{z}=\bigg(\frac{\partial v_{y}}{\partial x} - \frac{\partial v_{x}}{\partial y}\bigg)=\omega_{z} ~~~~~[vertical~~vorticity]
 \label{curlvz}
\end{equation}

\begin{equation}
 \centering
 (\bm{\nabla}.\bm{v})_{h}=\bigg(\frac{\partial v_{x}}{\partial x} + \frac{\partial v_{y}}{\partial y}\bigg)= d_{h} ~~~~~[horizontal~~divergence]
 \label{divvh}
\end{equation}

where $v_{x}$ and $v_{y}$ are the horizontal velocities and are computed by applying Fourier Local Correlation Tracking \citep[FLCT;][]{2004ApJ...610.1148W} on the Doppler velocities obtained from HMI. The granular structures, which appear as upflows and downflows in the Doppler velocity maps, are used to track the horizontal motions on the Sun. For the application of FLCT procedure, we have used parameters such as the Gaussian window ($\sigma=15$ pixels) and the time difference between the images as ($\bigtriangleup t \approx2$~min), similar to \citet{2016ApJ...824..120S}. Moreover, we have removed the {\it p} and {\it f} mode signals from the data before computing the velocities. This has been done by using a Gaussian-tapered filter that filters out all the signals above 1.2~mHz frequency. The nature of these physical quantities provides us with information on the fluid properties at the photosphere.

\begin{figure*}[ht]  
\centering
\includegraphics[width=0.5\textwidth]{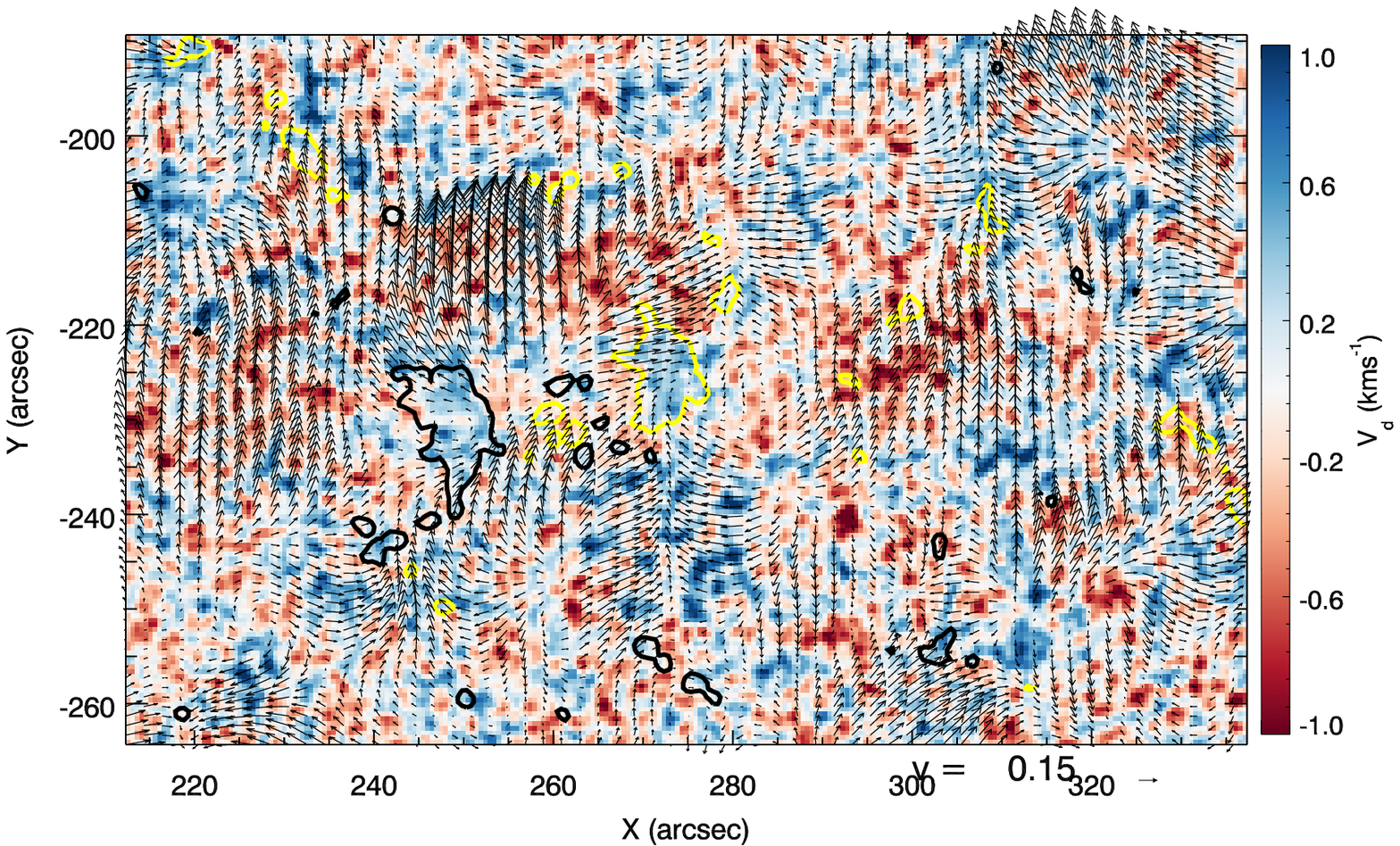}
\includegraphics[width=0.5\textwidth]{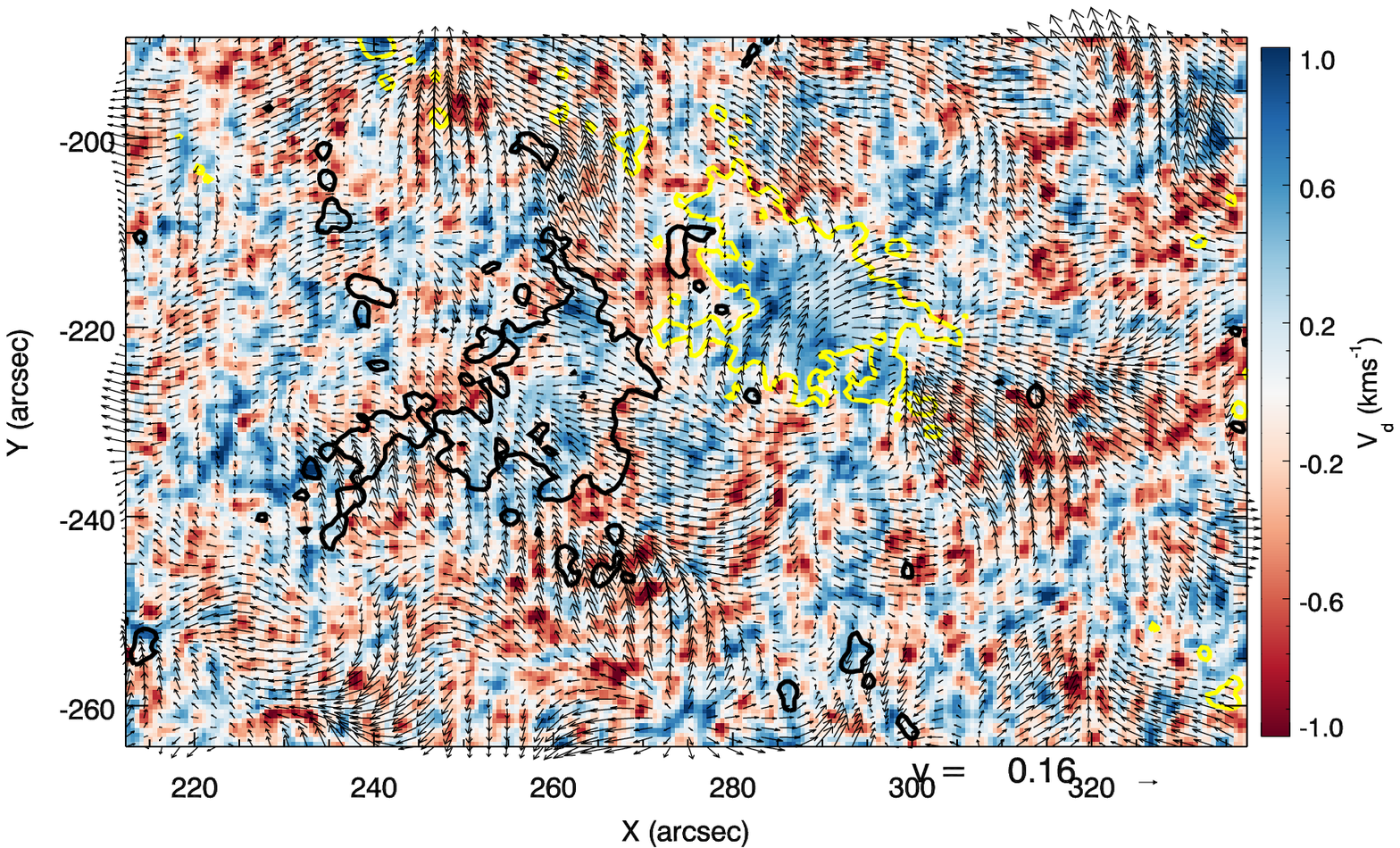}
\includegraphics[width=0.5\textwidth]{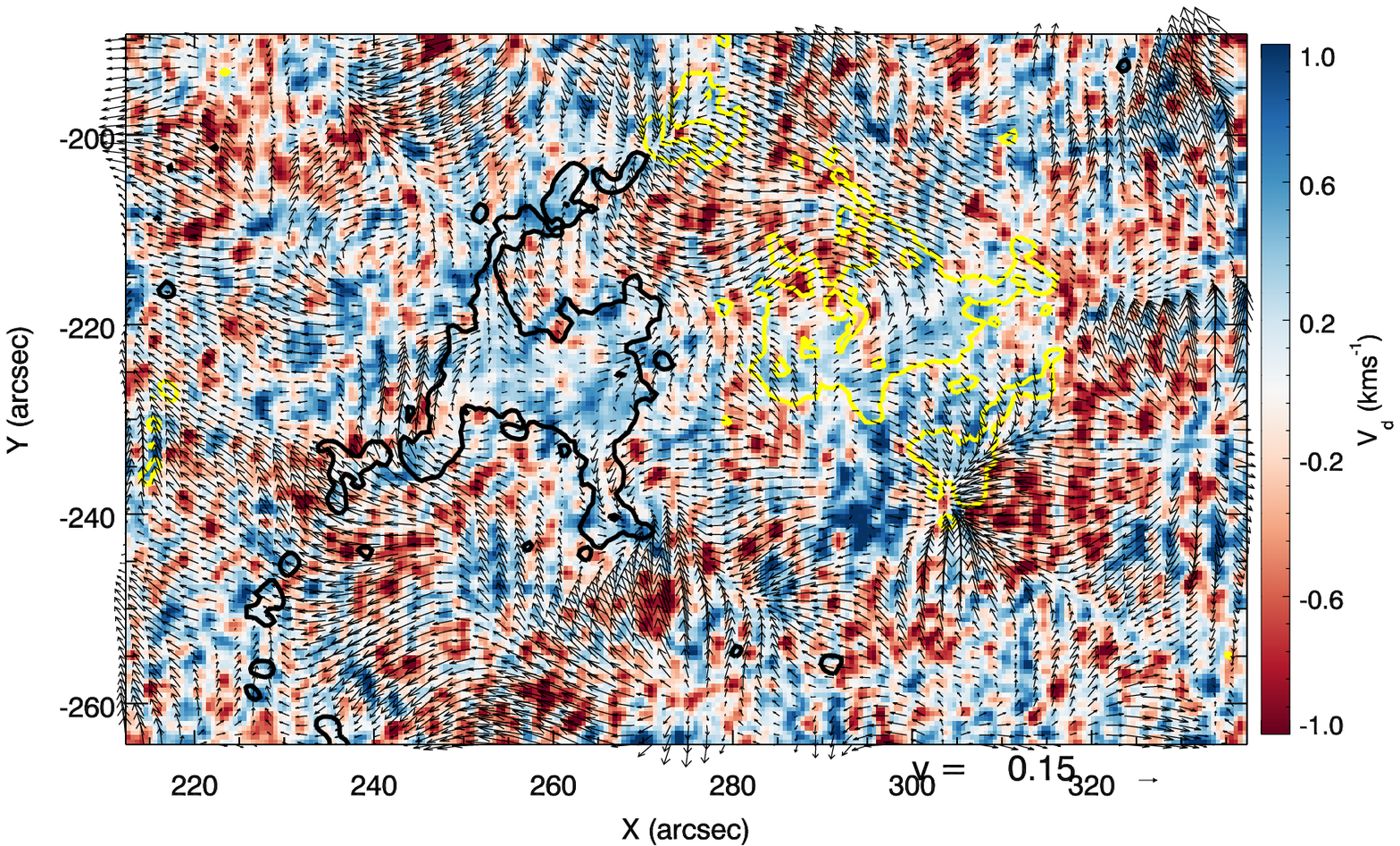}
\caption{Snapshots of the Doppler velocities corresponding to three of observation sets namely T1, T3, and T4. Over-plotted arrows are the horizontal motions tracked from fluid motions using FLCT. The over-plotted contours correspond to negative (black) and positive (yellow) magnetic field of $\pm$50~G.}\label{v_arrow}
\end{figure*}

In Fig.~\ref{v_arrow}, we show two snapshots of Doppler velocity maps. The over-plotted black arrows represent the horizontal motions tracked from the Doppler velocity using the FLCT procedure. The length of the arrow indicates the flow speed. We have over-plotted the contours of $\pm$50~G positive (magenta) and negative (black) magnetic field. These plots clearly show the swirly pattern in the plasma. It is more evident in the positive field region than in the negative magnetic field region. It is to be noted that not just the swirly motions associated with fluids results in vorticity but also shearing motions in plasma. The horizontal velocities computed here are then used to compute $\omega_{z}$ and $d_{h}$ for all the five datasets separately and are shown in Figs.~\ref{div_all} and ~\ref{cvz_all}. We emphasise that these quantities are derived at each time step. However, for the purpose of displaying, a time average for each set is plotted in corresponding panels.

\section{Results}\label{dar}

\begin{figure*}[ht]  
\centering
\includegraphics[width=0.8\textwidth]{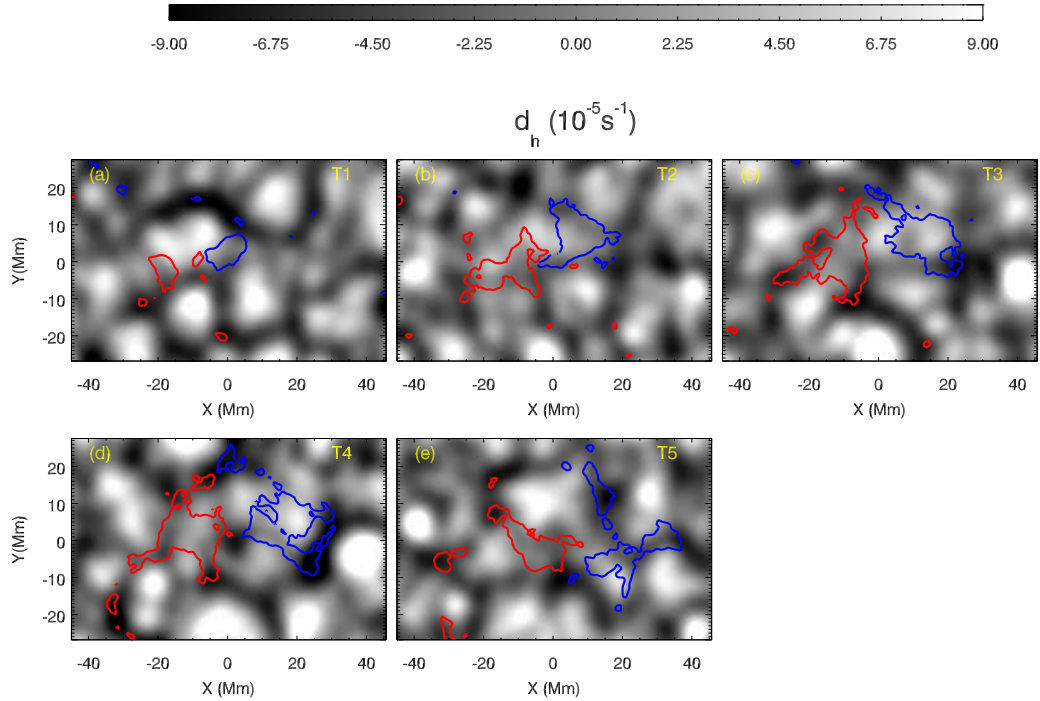}
\caption{Time averaged horizontal divergence maps corresponding to the five observation set namely T1, T2, T3, T4 and T5. The over-plotted contours correspond to negative (red) and positive (blue) magnetic field of $\pm$50~G.}\label{div_all}
\end{figure*}

Figure \ref{div_all} displays the time-averaged horizontal divergence maps corresponding to the five sets of observations. The bright (dark) regions depict the positive (negative) horizontal divergence representing upward (downward) plasma motion. The over-plotted contours correspond to negative (red) and positive (blue) magnetic flux density values of $\pm$50~G. As can be inferred from these maps, the flux emerges within the regions with negative divergence as well as at the edges of the areas with positive divergences. Later during evolution, it spreads in the areas with positive divergence (panels b, c and d).  The horizontal diverging flows push the emerging flux to the boundaries between the positive and negative divergence regions due to flux expulsion \citep{1982RPPh...45.1317P}.

Time-averaged maps of vorticities obtained for the five observation sequences are shown in Fig. \ref{cvz_all}, as labelled. The positive (negative) $\omega_{z}$ values represent the motion of the fluid in anti-clockwise (clockwise) direction. The over-plotted contours are the negative (red) and positive (blue) contours of $B_{LOS}$ similar to Figure~\ref{div_all}. The panel (a) of Figure \ref{cvz_all} distinctly shows that the two polarities of the emerging flux have opposite signs of vertical vorticity. The presence of opposite sign of vorticity is suggestive of the fact that the fluid at the two polarities is swirling in opposite directions. With passing time, this pattern changes, and we observe a mixed sense of vorticities in subsequent observation sets. However, the existence of differently directed swirly motions at both the polarities is still observed. We further note that towards the end of the observation, the direction of these swirly motions at two polarities are opposite to that seen at the beginning of the observations (see panel a and e).

\subsection{Influence of emerging flux on the fluid vorticity inside emerging magnetic field region}
\label{mregion} 

In order to have further quantitative measurements of vorticities and its time evolution at the two polarities of the emerging flux, we plot in the top panel of Fig.~\ref{caBtime_all} the spatially averaged signed vorticities for positive and negative magnetic flux for all the five sets of observations. In the bottom panel, we plot the 
evolution of the unsigned vertical vorticity. Note that for creating these plots we have only considered the pixels with magnetic flux density larger than $\pm$~10G, due to associated errors. It has to noted that we are using magnetic fields only to separate magnetic and non-magnetic regions. We have not used magnetic field to track velocities. Hence, a lower threshold of 10~G which of the order of the error in magnetic field measurements was used. Higher threshold in magnetic field could affect the results of non-magnetic regions.  In the top panel, the black curve represents the vertical vorticity for positive magnetic flux regions (leading polarity), whereas the red curve shows that for negative flux regions (trailing polarity). The figure reveals that the leading polarity primarily shows negative vorticity, which is according to the hemispheric rule, whereas the trailing polarity does not. Therefore, it is plausible to conclude that the two polarities are swirling in opposite directions, indicating either unwinding or winding of emerging flux tube. Similar to what was deduced from panel 'a' in Fig.~\ref{cvz_all}, the plots reveal that at the beginning of the flux emergence, the vertical vorticities at the two polarities are of opposite signs, suggesting rotation in the opposite direction. During the course of the evolution, though there are changes and the magnitude of vertical vorticity reduces, the sense of twist remain almost opposite, up to T4. In T5, we observe that the vorticity is mostly negative at both the footpoints, indicating that the fluid is going back to the original state, which is typical for the southern hemisphere of the Sun. Moreover, all these changes observed are more than the standard error estimates which lies between 5{--}9${\bm \times 10^{-7}\ s^{-1} }$.

We add a caveat here on the breaks in vorticity observed in these time evolution plots. When horizontal velocities are derived from the Doppler velocities using FLCT, at the edges, {\it i.e.,} at the start and end of the time series, we get incorrect velocity measurements. Hence, we remove velocity measured at the edges to derive vorticity and divergence. This creates a gap of 15~minutes between each dataset. Additionally, to remove the small-scale evolving features, we have performed one hour of smoothening of the derived vorticity. This introduces a gap by additional two hours in the data. So a total of 2.5~hours between the datasets. The magnetic field evolution shown in Fig.~\ref{mg_flux} also has the same time gap as the vorticity and divergence to match the results observed. 

From the bottom panel, we find that in the initial stages of flux emergence (T1), the vertical vorticity is fluctuating around 3.5 $ \times 10^{-5}\ s^{-1} $. However, from the start of the 2nd set of observation (i.e., during the strong emergence of magnetic flux) till the end (T5) the unsigned vertical vorticity shows a monotonic increase. We note that precisely at the same time when the signed vertical vorticity is found to decrease. This result could primarily be attributed to the generation of more oppositely directed vorticities in the fluids during the strong emergence of the flux.

\subsection{Influence of emerging flux on the fluid vorticity in the surrounding non-magnetic regions}\label{vortrad}

We further study the effects of emerging flux to the surrounding non-magnetic regions. For this purpose, we have defined the non-magnetic areas to be regions with magnetic flux density less than 10~G. We perform a similar analysis as was done for magnetic regions in \S\ref{mregion}. In Fig.~\ref{ca0Gtime_all}, we plot the spatially signed averaged (top panel) and unsigned averaged vertical vorticity (bottom panel) as a function of time for all the five sets of observations. While the signed averaged curve shows a significant fluctuation in the vertical vorticity that is highest at the peak of the emerging flux, the unsigned average plot shows a monotonic increase in the vertical vorticities in the non-magnetic region. From the top panel, we note that most of the time, the negative vorticity is more dominant over the positive, whereas at the time of intense flux emergence, i.e., during T3 and T4, the positive vorticity is dominant. The bottom panel shows a monotonic increase in the vorticity, similar to that is observed for the magnetic regions. It is interesting to note that the magnitude of the vorticity in magnetic and non-magnetic regions are almost the same. We further note that the unsigned averaged vorticity, for both magnetic (bottom panel of Fig.~\ref{caBtime_all}) and non-magnetic regions (bottom panel of Fig.~\ref{ca0Gtime_all}), keeps increasing even after the flux emergence has come to a halt.

The next obvious step is to study the extent to which the emerging flux region affects the fluid motion in the quiet Sun (B$_{LOS}<$10G). For this purpose, we drew concentric circles around the emerging flux as shown in the top panel of Fig.~\ref{mg08cir} and identify them as R1{--}R6 as labelled. The first circle has a radius of 25$\arcsec$ and the consecutive circles increase in radius by 25$\arcsec$. Since close to the equator the vorticity values tend to reduce towards zero, we have considered only up to about 60$\arcsec$ in the southern hemisphere for this analysis. In the bottom panel of Fig.~\ref{mg08cir}, we plot the evolution of the signed averaged vorticities within the concentric cells for all the five sets of observations for the non-magnetic region. As can be inferred from the plots, the vorticity in the region R1 {\it i.e.}, very close to the emerging flux region, the fluctuations in the vorticity is most substantial. These fluctuations reduce in the outer cells. This is suggestive of the fact that the effects of emerging flux on the fluid motion in the surrounding non-magnetic region is strongest in the close vicinity and gradually reduces as we further move out.

\subsection{Relationship between fluid's vertical vorticity and horizontal divergence}\label{vor_div}

There have been several studies showing a linear correlation between vertical vorticity ($\omega_{z}$) and horizontal divergence ($d_{h}$) for quiet Sun region that has been attributed to Coriolis effects \citep[e.g.,][]{2000SoPh..192..177D,2003ESASP.517...43G, 2016ApJ...824..120S}. However, such correlation gets 
altered in magnetic regions \citep[][]{2016ApJ...824..120S}. Therefore, it is imperative to study the correlation between these two quantities in the emerging flux regions and their evolution. For that purpose, we have plotted in Fig.~\ref{cadiv0G_all} the time-averaged vertical vorticity as a function of horizontal divergence obtained for five difference sets of observations as labelled. For this purpose, first, we have identified all the quiet-Sun pixels in all the five sets of observations with $|B_{LOS}|<$10~G. For all these pixels, we have binned the $\omega_{z}$ within a bin size 20~$\mu s^{-1}$ for divergence and plotted the variation of $\omega_{z}$ and $d_{h}$ for each observation set. For comparison with quiet Sun, we have reproduced the vertical vorticity as a function of horizontal divergence curve (in red) for a quiet Sun region from \citet{2016ApJ...824..120S}. The figure reveals that the solid black line corresponding to the first set of observation T1 (right at the start of 
flux emergence) shows the same behaviour of linear correlation as that of the solid red line. However, this linear correlation starts to get altered for the observation set T2 that continues with emerging flux. The curve obtained for the last set, i.e. T5, shows behaviour similar to that obtained for the northern hemisphere, i.e., for negative divergence vorticity is positive and vice-versa \citep{2016ApJ...824..120S,2000SoPh..192..177D, 2003ESASP.517...43G}.

\subsection{Energetics}

\begin{table}[ht]
\centering
\caption{\footnotesize Maximum and averaged magnetic and kinetic energy densities for the five time intervals. All the energy densities are in erg cm$^{-3}$. }
\begin{tabular}{ccccc}
\hline
{\bf Time Interval}& ${\bm E_{B}^{max}}$ & ${\bm E_{B}^{avg}}$ & ${\bm E_{K}^{max}}$ & ${\bm E_{K}^{avg}}$ \\[0.5ex]
 & \multicolumn{4}{c}{${\bm erg~cm^{-3} }$} \\[0.5ex]
\hline
T1 & $7.45\times 10^{4}$ & $1.27\times 10^{3}$ & $2.83\times 10^{3}$ & $57.3$ \\[0.5ex]
T2 & $7.87\times 10^{4}$ & $1.94\times 10^{3}$ & $1.87\times 10^{3}$ & $27$ \\[0.5ex]
T3 & $8.2\times 10^{4}$ & $2.04\times 10^{3}$ & $2.8\times 10^{3}$ & $58.9$ \\[0.5ex]
T4 & $6.44\times 10^{4}$ & $1.61\times 10^{3}$ & $2.98\times 10^{3}$ & $58.9$ \\[0.5ex]
T5 & $2.4\times 10^{4}$ & $9.28\times 10^{2}$ & $4.22\times 10^{3}$ & $90.11$ \\[0.5ex]
\hline
\end{tabular}
\label{tab:energy}
\end{table}

In order to understand the involved energetics in the emerging flux region, we compute the magnetic ($B^{2}/(8\pi)$) as well as the kinetic energy densities ($\rho v^{2}/2$) during the emergence, where, B is the magnetic field, $\rho$ is mass density that is taken from FAL-93 \citep{1993ApJ...406..319F} Model-C as $=\ 2.75\times 10^{-7}$~g~cm$^{-3}$. v is the horizontal velocity that is obtained as $\sqrt{v_x^2 + v_y^2}$, which are computed using FLCT (as shown in Fig.~\ref{v_arrow}). 

Table~\ref{tab:energy} lists the estimates of maximum magnetic energy density $E_{B}^{max}$ (column 2) and maximum kinetic energy density $E_{K}^{max}$ (column 4) obtained in each set of observation. Column 3 and 5 are the spatially and temporally averaged magnetic and kinetic energies. We note that for this analysis we have only used pixels with magnetic flux density $B > \pm 50~G$. Regions with high magnetic fields were considered to avoid the small-scale field, which may or may not be related to the emerging flux region. 

The analysis shows that the magnetic energy density is always higher than the kinetic energy density for all the sets of observations. Moreover, the magnetic energy increases with time until the third set of observations and decreases thereafter. This finding is consistent with the magnetic flux as seen in Figure~\ref{mg_flux}. We further note that, while the magnetic energy increases, kinetic energy slightly decreases and vice-versa in the later phase. The presence of higher magnetic energy at all times makes it plausible to conclude that the magnetic field has sufficient energy to alter surrounding fluid motions. 

We would like to point out that, in an ideal situation, for a flux tube surrounded by a completely field-free plasma, there is no tangential force on the flux tube surface i.e., the magnetic pressure force is perpendicular to the flux surface and the tangential tension force would not have any effect on the surrounding plasma. But in reality a flux tube in a stratified plasma is subject to instabilities like kink or sausage instabilities \citep{2014masu.book.....P}. These would hence lead to tension forces tangential to the field. This can cause exchange of vortical motions that are of magnetic in origin, viz. the magnetic baroclinicity and tension terms of the vorticity evolution equation in MHD (see, e.g. Blandford and Thorne, Lecture Notes on Applications of Classical Physics)\footnote{http://www.pmaweb.caltech.edu/Courses/ph136/yr2011/}. In this work, we have measured and studied motions around an emerging flux region and have attempted to link their vortical nature to the above sources of magnetic origin. The details of how magnetic field generates vorticity is not aimed observationally in our work. 

\section{Summary and Conclusion}\label{SandC}

In this paper, we have studied the evolution of an emerging bipolar region from its birth using the observations recorded with HMI. For this purpose we have computed the vertical vorticity and horizontal divergences by tracking the horizontal motions using the Fourier Local Correlation Tracking methods \citep[FLCT,][]{2004ApJ...610.1148W}. We summarise the results below.

\begin{figure*}[ht]  
\centering
\includegraphics[width=0.8\textwidth]{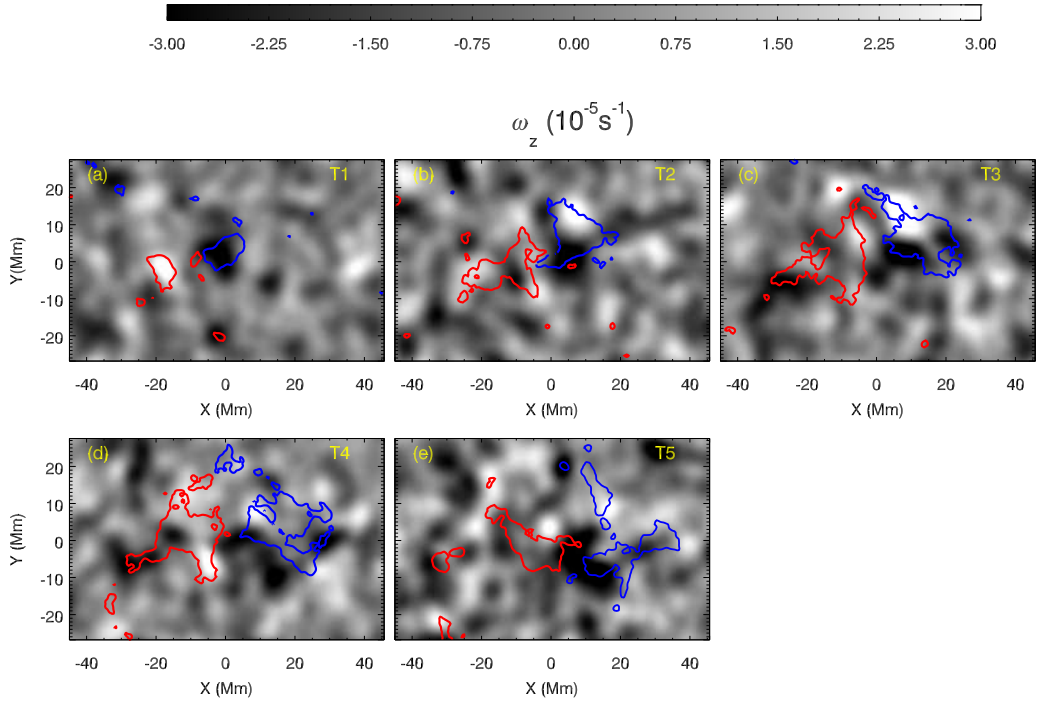}
\caption{Same as Fig.~\ref{div_all} but for vertical vorticity.}\label{cvz_all}
\end{figure*}

\begin{figure*}[ht]  
\centering
\includegraphics[width=0.85\textwidth]{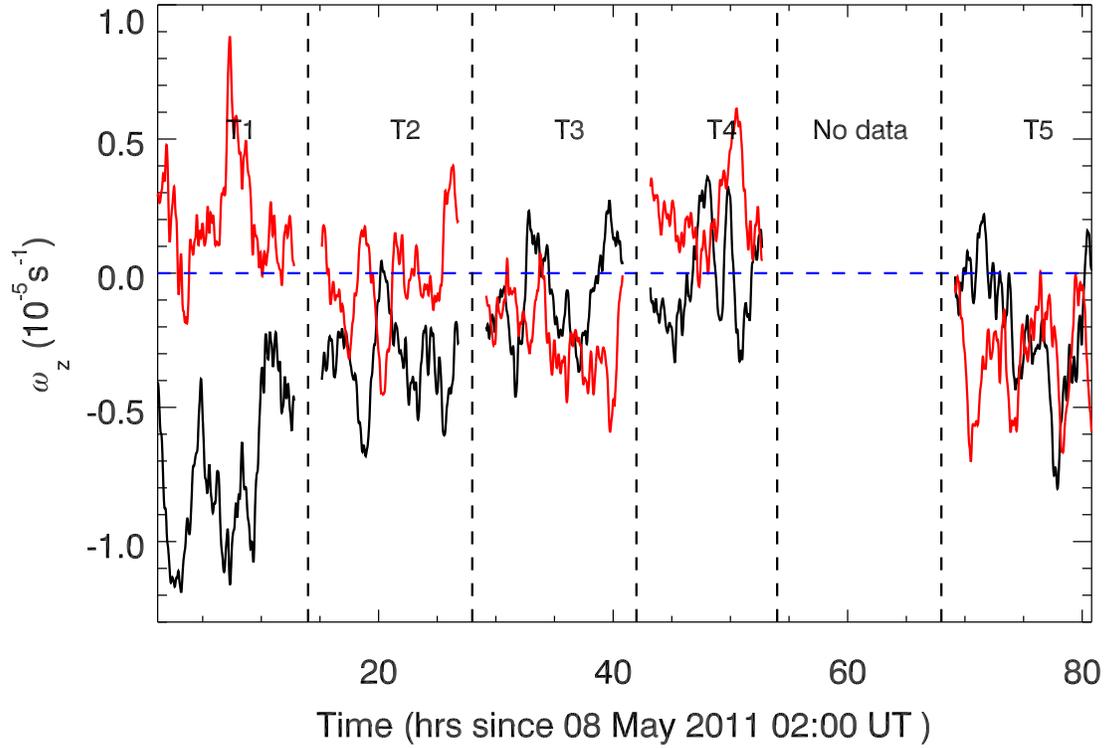}
\includegraphics[width=0.81\textwidth]{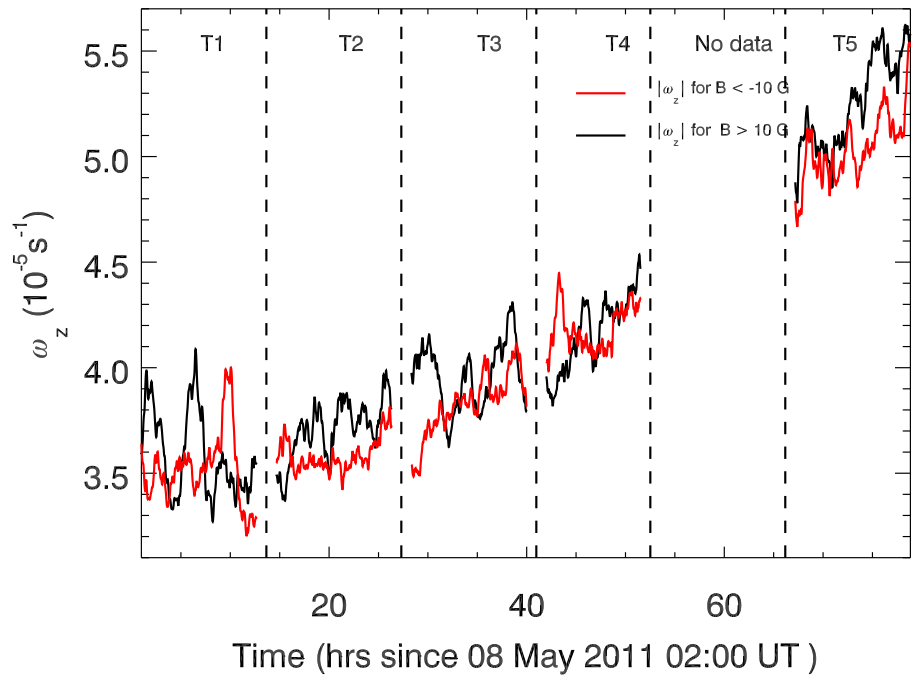}
\caption{Top panel: Evolution of spatially signed averaged vertical vorticity for five different sets of observations as labelled for magnetic regions ($|B_{LOS}|>10$~G). The black and red curves show the evolution of signed vorticity for positive and negative magnetic flux, respectively. Bottom panel: Evolution of unsigned averaged vertical vorticity.} \label{caBtime_all}
\end{figure*}

\begin{figure*}[ht]  
\centering
\includegraphics[width=0.85\textwidth]{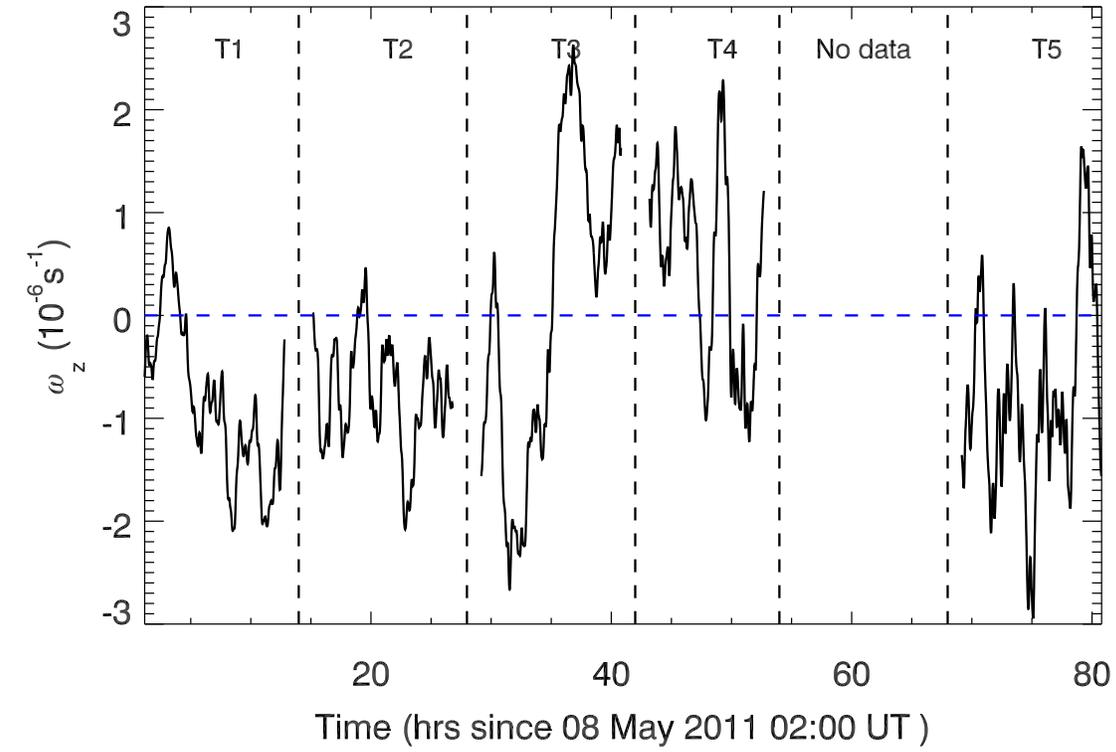}
\includegraphics[width=0.85\textwidth]{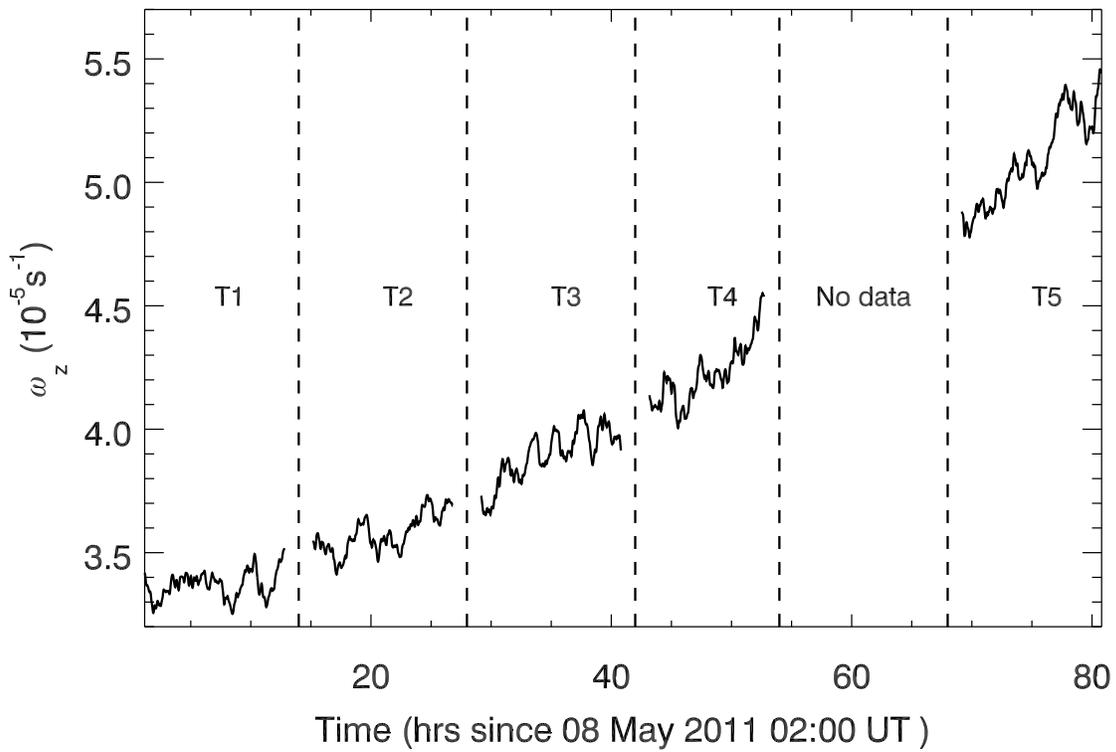}
\caption{Same as Figure~\ref{caBtime_all} but for non-magnetic regions ($|B_{LOS}|<10$~G). }\label{ca0Gtime_all}
\end{figure*}

\begin{figure*}[ht]  
\centering
\includegraphics[width=0.8\textwidth]{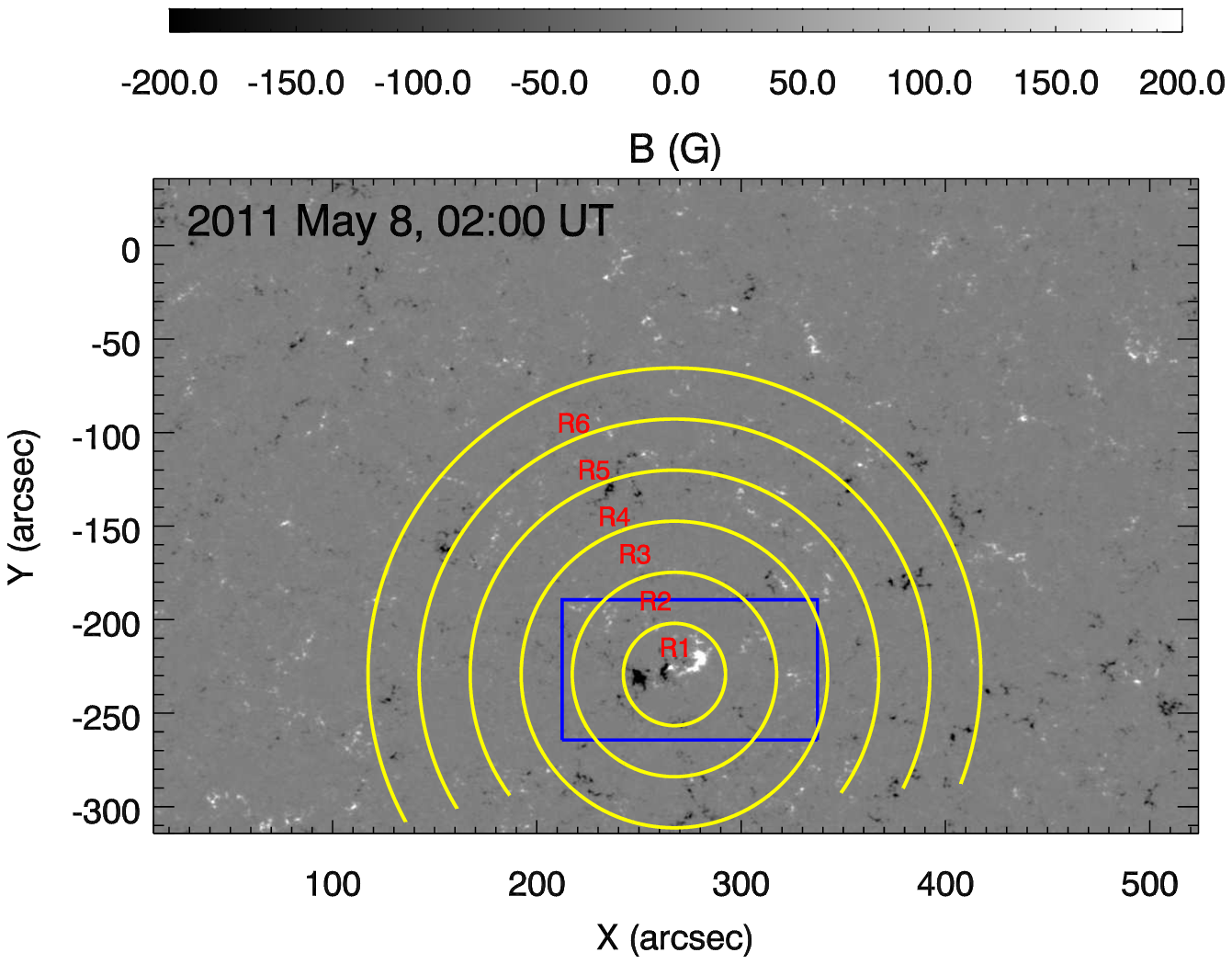}
\includegraphics[width=0.85\textwidth]{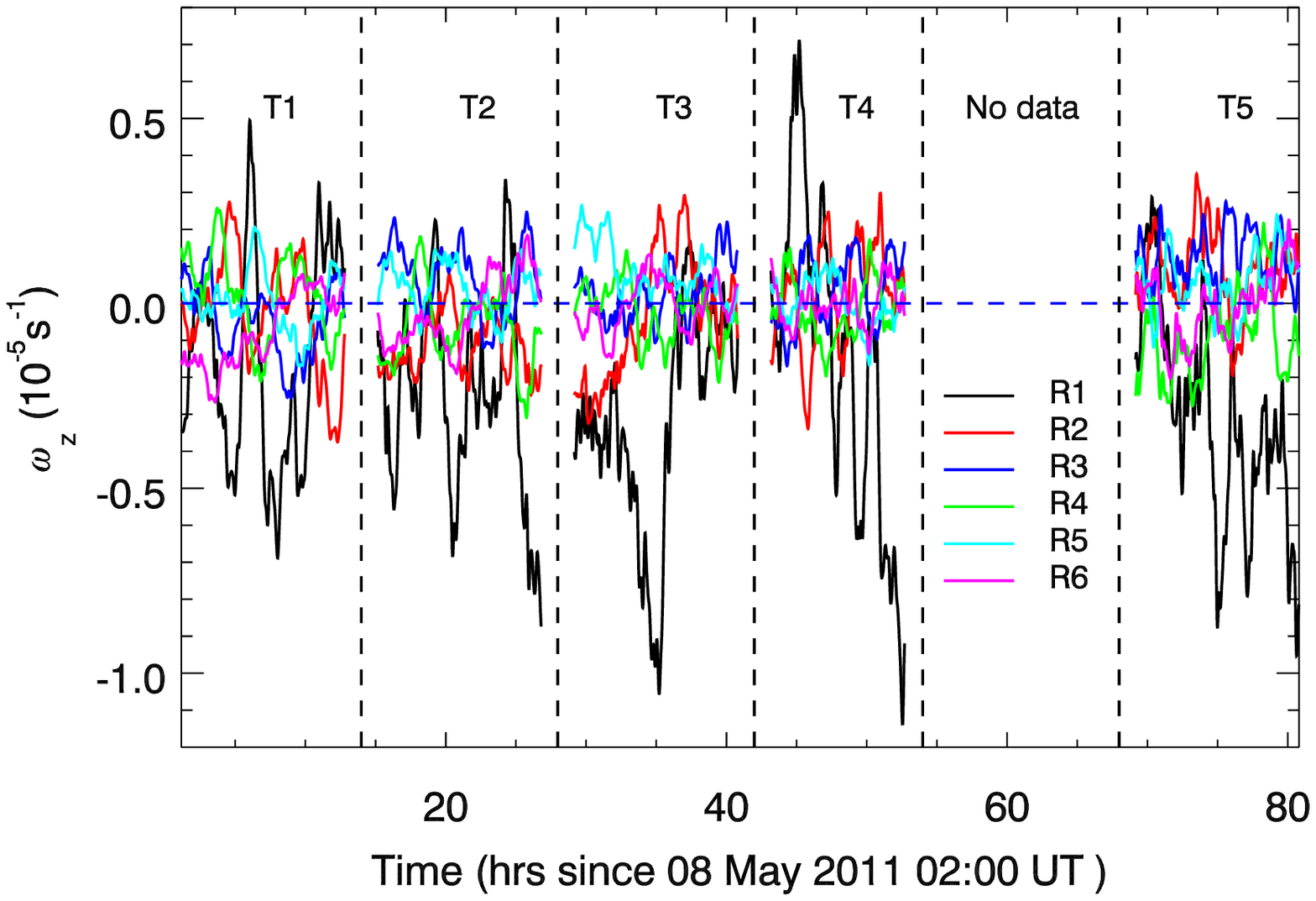}
\caption{Top panel: Same as Fig.~\ref{mg08} but over-plotted concentric circles marked R1{--}R6, representing the regions used for the analysis of the vorticity evolution in radial directions. Bottom panel: Signed averages of the vorticity computed within the concentric cells as labelled.}
\label{mg08cir}
\end{figure*}

\begin{figure*}[ht]  
\centering
\includegraphics[width=0.9\textwidth]{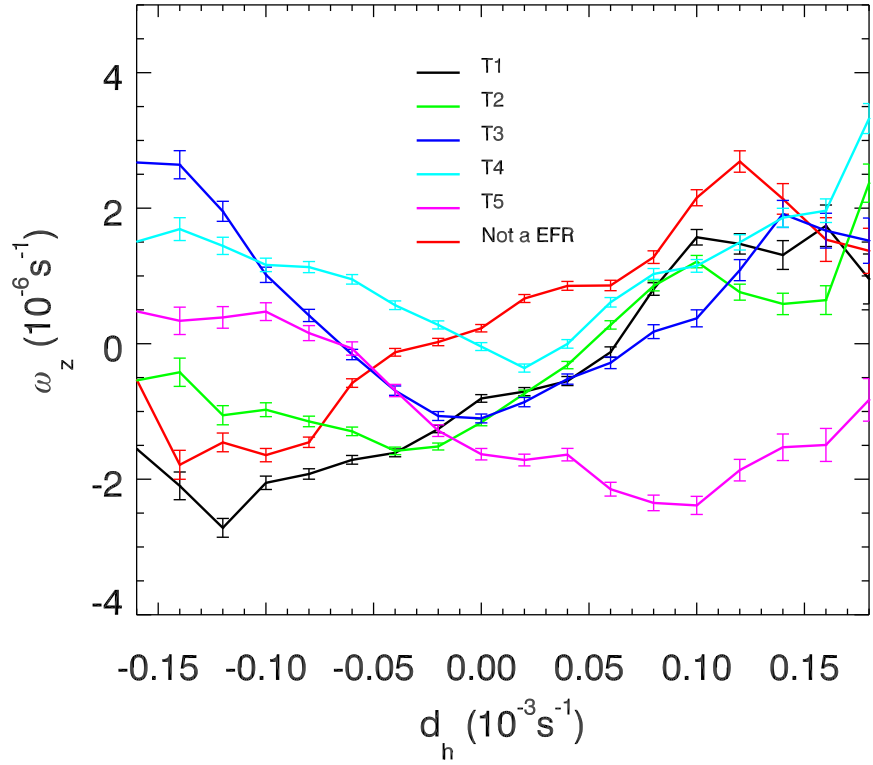}
\caption{Time averaged vertical vorticities as a function of horizontal divergences for non-magnetic regions  ($B_{LOS}<10$~G) as labelled. For comparison we have plotted in red the variation of time averaged vorticity as a function of horizontal divergence for a completely different quiet-Sun region taken at the same latitude and longitude taken from \cite{2016ApJ...824..120S}.}
\label{cadiv0G_all}
\end{figure*}

\begin{enumerate}

\item At the initial phase of flux emergence, the two bipolar footpoints of emerging flux harbor {\bf swirly} motions or vorticities of opposite signs (see Figs.~\ref{cvz_all} \&~\ref{caBtime_all}). The sign of vorticity in the leading polarity confirms to what the hemispheric rule would predict for the southern hemisphere of the Sun (i.e., clockwise rotation).

\item The unsigned average of vorticity computed for the magnetic region show a monotonic increase with time, i.e. with emerging flux. The signed averages of vorticity show strong fluctuations in the early phases of flux emergence. In the non-magnetic regions the increase in vorticity is found to be strongly correlated with the emerging magnetic flux (see Fig.~\ref{caBtime_all} \&~\ref{ca0Gtime_all}).

\item We have further studied the spatial extent till which the emerging flux affects the fluid motions in the quiet-Sun. We find that the fluid close to the emerging flux region is most strongly affected. Such effect reduces as we further move away from the emerging flux region (see Fig.~\ref{mg08cir}).

\item Our study further reveals that right at the beginning of flux emergence, there is a linear correlation between vortical vorticity and horizontal divergence. However, during the emergence, this correlation gets altered and become opposite to what is observed for quiet-Sun in the southern hemisphere (see Fig.~\ref{cadiv0G_all}).

\item The magnetic energy density (both average and maximum) is found to be greater than the kinetic energy density at all times during the flux emergence (see Table~\ref{tab:energy}).

\end{enumerate}

In our analysis, we found that during the early phase of the flux emergence, the two footpoints of the magnetic field has twist in opposite directions, with only the leading polarity following clock-wise (negative vorticity) direction of the swirly motions typical for southern hemisphere \citep[][]{2016ApJ...824..120S} according to hemispheric helicity rules \citep{1990SoPh..125..219S,1995ApJ...440L.109P,1997SoPh..174..291A,1998ApJ...496L..43B,2014SSRv..186..285P}. Whereas, the trailing polarity twists in counter-clockwise (positive vorticity). Such disagreement in the vorticity (swirly motions) at the two footpoints can be attributed to a conclusive signature of the unwinding of the twisted flux tube. It has to be noted that the opposite direction of swirl at the two footpoints can either wind or unwind the flux tube. But since the signed vorticity were reducing over time, we have concluded that the magnetic flux tubes are unwinding. There could also be other mechanisms for example, the plasma could be draining down the helical field lines resulting in spiral-like structure with opposite vorticity or it could even occur due to a simple rotation of a flux tube as a whole without any change in it's internal twist. But again, this would lead to the main question why is it that the footpoints have opposite vorticity observed. We further note that observed opposite direction of twist is for about 1{--}2 days during the emergence, which is similar to the observed helicity injection rate observed in corona during the time of flux emergence \citep[see, e.g.,][]{2003ApJ...593.1217P}. The observation of helicity injection in the corona during flux emergence has been attributed to torsional Alfv\'en waves generated due to the rotation of the emerging flux tubes \citep{2003ApJ...593.1217P}.

Another important fact is that, we observe certain periodic variations in the vorticity with a period between 5{--}10 hours derived from the Doppler velocities. These periodic nature observed in vorticity alone i.e., these are not observed in the Doppler velocity data. Hence, we can rule out the fact that these are SDO orbital periodic variations observed in the data \citep{2012SoPh..279..295L}. These periodic behaviour observed in vortical motions are real and the origin of these is unknown. 

Our results also display the linear correlation between the vertical vorticity ($\omega_{z}$) and the horizontal divergence ($d_{h}$) during the very initial stage of the flux emergence. This result is similar to that is observed for quiet Sun using time-distance helioseismology \citep{2000SoPh..192..177D,2003ESASP.517...43G} as well 
as FLCT \citep[][]{2016ApJ...824..120S}. \citet{2016ApJ...824..120S} find that this behaviour is only valid for non-magnetic regions and the correlation is altered for magnetic regions. This is opposite to what is observed in our study during the emerging flux. This discrepancy could be attributed to the unwinding of the twisted flux tubes. In that scenario, the untwisting flux tube may impart its twist to the surrounding fluids. This scenario is also justified from the fact that at all time during the flux emergence, the magnetic energy density is at least ten times greater than the kinetic energy density.

The observations of twisting and untwisting of flux tubes that may generate the torsional Alfv$\acute{\text{e}}$n waves are also important to address the problem of the heating of the solar atmosphere \citep{2016NatPh..12..179J,2018MNRAS.480.2839L}. To the best of our knowledge, this is the first study reporting the unwinding of flux tube during the flux emergence. Further work involving a statistical study is required to comprehend such phenomena fully. \\ \\

C.R.S. and D.T. acknowledge support from the Max-Planck India Partner Group of MPS on “ Coupling and Dynamics of the Solar Atmosphere ” at IUCAA. The partner group is funded by MPG and DST(IGSTC), Government of India. A part of the data-intensive numerical computations required in this work were carried out using the High Performance Computing facility of the Indian Institute of Astrophysics, Bangalore and the rest using computational facility in IUCAA. This work has utilized extensively the HMI/SDO data pipeline at the Joint Science Operations Center (JSOC), Stanford University. We also thank Prof. P. Venkatakrishnan for his inputs on the results of this paper which helped us to improve the manuscript. We thank the referee for the insightful comments which has improve the paper.



\end{document}